\documentclass[useAMS,usenatbib,usegraphicx]{mn2e}
\usepackage{times, lscape}

\title[6 $-$ 200 $\mu$m emission of LDN~1688]
{Observations of 6 $-$ 200 $\mu$m
emission of the Ophiuchus cloud LDN~1688\thanks{Based
on observations made with {\it ISO}, an ESA project with instruments funded by ESA member
states (especially the PI countries: France, Germany, the Netherlands and the United
Kingdom) and with the participation of ISAS and NASA. 
{\it Herschel} is an ESA space observatory with science instruments provided by European-led Principal Investigator consortia and with important participation from NASA.}}

\author[M. G. Rawlings et al.]{M. G. Rawlings$^{1,2,3,4}$
\thanks{E-mail: mrawling@nrao.edu},
M. Juvela$^{4,6}$, K. Lehtinen$^{4,6}$, K. Mattila$^{4}$ \& D. Lemke$^{5}$\\
$^{1}$National Radio Astronomy Observatory, 520 Edgemont Road, Charlottesville, VA 22903, U. S. A.\\
$^{2}$Joint ALMA Observatory / European Southern Observatory, Alonso de C\'{o}rdova 3107, Vitacura 763-0355, Santiago, Chile\\
$^{3}$Joint Astronomy Centre, 660 N. A$`$ohoku Place, Hilo, HI 96720, U. S. A.\\
$^{4}$Observatory\thanks{Closed down on 31 December 2009}, University of Helsinki, FI-00014 Helsinki\\
$^{5}$Max-Planck-Institut f\"ur Astronomie, K\"onigstuhl 17, Heidelberg, D-69117, Germany\\
$^{6}$Department of Physics, University of Helsinki, FI-00014 Helsinki}

\begin{document}

\date{Accepted 2012 October 16. Received 2012 October 16; in original form 2011 November 8}

\pagerange{\pageref{firstpage}--\pageref{lastpage}} \pubyear{2012}

\maketitle

\label{firstpage}

\begin{abstract}

We examine two positions, ON1 and ON2, within the Ophiuchus cloud
LDN~1688 using observations made with the ISOPHOT instrument aboard
the ISO satellite. The data include mid-infrared spectra
($\sim$6--12\,$\mu$m) and several photometric bands up to 200\,$\mu$m.
The data probe the emission from molecular PAH-type species,
transiently-heated Very Small Grains (VSGs), and large classical dust
grains.

We compare the observations to earlier studies, especially those
carried out towards an isolated translucent cloud in Chamaeleon (Paper
I). The spectra towards the two LDN~1688 positions are very similar to
each other, in spite of position ON1 having a larger column density
and probably being subjected to a stronger radiation field. The ratios
of the mid-infrared features are similar to those found in other
diffuse and translucent clouds. Compared to paper I, the
7.7/11.3\,$\mu$m band ratios are lower, $\sim$2.0, at both LDN~1688
positions. A continuum is detected in the $\sim$10\,$\mu$m region. This
is stronger towards the position ON1 but still lower than on any of
the sightlines in Paper~I. The far-infrared opacities are higher than
for diffuse medium. The value of the position ON2,
$\tau_{200}/N(H)=3.9\times 10^{-25}\,{\rm cm}^2/H$, is twice the value
found for ON1.

The radiation field of LDN~1688 is dominated by the two embedded B
type double stars, $\rho$ Oph AB and HD 147889, with an additional
contribution from the Upper Sco OB association. The strong heating is
reflected in the high colour temperature, $\sim$24\,K, of the large
grain emission. Radiative transfer modelling confirms a high level of
the radiation field and points to an increased abundance of PAH
grains. However, when the hardening of the radiation field caused by
the local B-stars is taken into account, the observations can be
fitted with almost no change to the standard dust models. However, all
the examined models underestimate the level of the mid-infrared
continuum.
\end{abstract}

\begin{keywords}
Infrared: ISM -- ISM: clouds -- ISM: molecules -- dust, extinction

\end{keywords}

\section{Introduction}

\subsection{Background}

The mid- to far-infrared (MIR, FIR) emission spectrum of the diffuse
interstellar medium (ISM) and interstellar clouds has frequently been described via a three-component IR dust model (e.g. \citealt{puget}).
Such models typically feature a mixture of (i) large aromatic organic ions or molecules producing the so-called Unidentified Infrared Bands (UIBs or UIR bands, also termed Infrared Emission Features, or IEFs) (\citealt{bakes2001}
and references therein), (ii) a population of transiently-heated Very Small Grains, or VSGs, \citep{sellgren} generating mid- to
far-IR emission, and (iii) larger ($\sim 100 - 2000$ \AA ) ``classical'' dust grains in thermal equilibrium
emitting in the far-IR ($\lambda \geq 80 \mu$m; e.g. \citealt{mathis1996}, \citealt{li97}).

In \citet{rawlings2005} (hereafter Paper 1), we presented an analysis of the relative contributions of these three components for G~300.2~-16.8,
a local, isolated, high-latitude translucent cloud. Through a comparison of the relative contributions of the three components, it was
demonstrated that the energy requirements of both the observed emission and optical scattered light could be accounted for by the
incident local interstellar radiation field (ISRF) alone. 

At the same time that G~300.2~-16.8 was observed, comparable observations were taken of other Galactic IR
emission regions using the ISOPHOT instrument \citep{lemke96} aboard the {\em Infrared Space Observatory (ISO)}
 \citep{kessler}, to allow direct comparisons to be made between environments with strongly differing ambient ISRF strengths. One
such region was LDN 1688, the main molecular cloud of the $\rho$ Ophiuchi cloud complex. 
There have been relatively few studies of high radiation field environments such as LDN~1688 in which a single 
dataset has been used to compare the mid- to far- IR emission spectra.
Besides its wide wavelength coverage 3 -- 240  $\mu$m,  ISOPHOT also had a large number of filter bands covering
the mid- to far-IR wavelength range more uniformly than its successors (e.g. {\it Spitzer}, \citealt{werner2004} and {\it Herschel}, \citealt{pilbratt2010}), important 
for specifying the different dust components.

\subsection{LDN~1688 and its dust emission}

The $\rho$ Oph star-forming cloud complex has
been a target of many studies (for a review of the properties of the region, see e.g. \citealt{wilking2008}), 
and contains numerous newly formed stars and embedded YSOs \citep{bontemps1996}.
It is now known to contain a star-forming cluster with $\sim$100 stars with ages $<1$ Myr old \citep{padgett2008}.
Due to a combination of the presence of associated stars and large column densities of gas and dust, 
LDN 1688 exhibits strong multiple-component IR emission. 
Detailed analysis and modelling of IRAS observations of this region have been conducted
by \citet{bernard93}.

\begin{figure*}
\center
\includegraphics[angle=0,width=16cm]{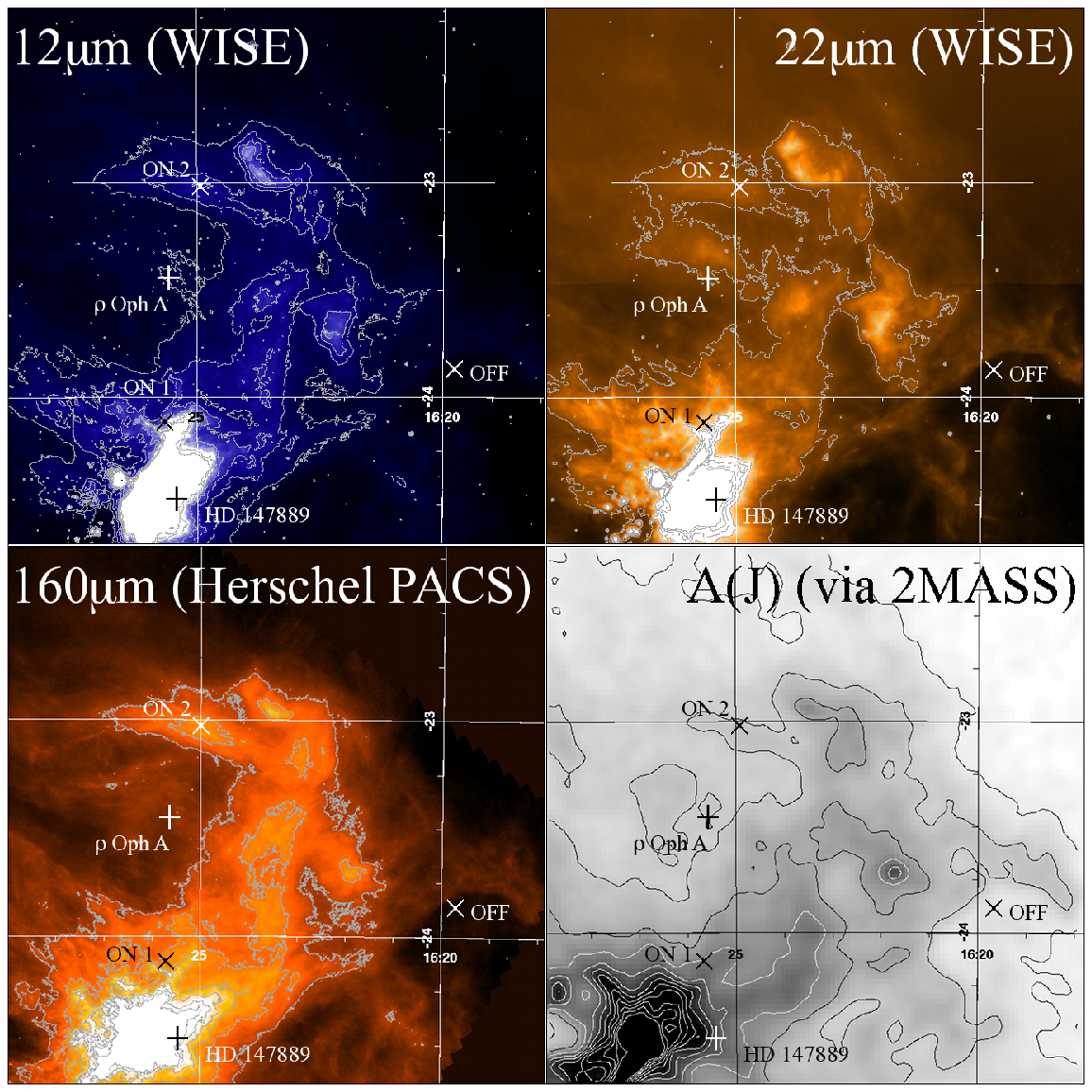}

 \caption{Infrared emission and visual extinction in LDN~1688. The upper two images are WISE \citep{wright2010} maps for the 12-$\mu$m and 22-$\mu$m filter bands. The lower left is the corresponding {\it Herschel}  PACS \citep{poglitsch2010} Red map, centred on 160$\mu$m. These three maps reflect the distribution of the PAHs, VSGs and big ``classical'' grains, respectively. The lower right image is a $A(J)$ map obtained using the 2MASS $JHK_{S}$ colour excesses of stars visible through the LDN~1688 central region via the multi-band colour excess method. The pixel size for the $A(J)$ map is 1.4' with a Gaussian FWHM = 4' used as a smoothing function for the individual extinction values. All frames shown are 2.5$^{\circ}$ squared, with $\alpha , \delta (J2000.0)$ co-ordinates shown. The positions observed by us using the ISOPHOT instrument aboard the ISO spacecraft are indicated by the symbol ``$\times$'', where ``OFF" is the off-source reference position and ON1 and ON2 are the on-source positions. The bright local stars providing a significant fraction of the local ISRF are denoted by ``$+$'' symbols. The contours for the 12-$\mu$m filter are as follows: $1000-1600$ data numbers (DN) in steps of 100. The contours for the 22-$\mu$m filter are as follows: $290-340$ DN in steps of 10. The contours for the 160-$\mu$m filter are as follows: $0.2-1.2$ Jy / pixel in steps of 0.2. The contours for the $A(J)$ map are as follows: $0.4-5.9$ magnitudes in steps of 0.5. Note the close spatial correlation between the large grains and the extinction map.}

\label{fig: WISE_Herschel_AJ_ims}
\end{figure*}

We present here ISOPHOT data between 6 and 240 $\mu$m, showing observations
of two ON-source positions in LDN 1688 and a nearby OFF position (see Fig. \ref{fig: WISE_Herschel_AJ_ims}).
In the spirit of the aforementioned three-component dust model, imaging at 12, 22 and 160 $\mu$m 
are shown in Fig. \ref{fig: WISE_Herschel_AJ_ims} to reflect the distributions of the PAH, VSG and large, 
classical grains, respectively. Position ON1 is close to the central brightness maximum of the cloud 
containing the embedded star cluster. This position was previously found to exhibit a
high {\it IRAS} $I_{12 \mu\rm{m}}$/$I_{100 \mu\rm{m}}$  ratio \citep{bernard93}.  
The second sightline, ON2, is located in a dense filament at the Northern edge of the
LDN~1688 cloud and is exposed to the external radiation from the Upper Scorpius OB association. 
It exhibited a low {\it IRAS} $I_{12 \mu\rm{m}}$/$I_{100 \mu\rm{m}}$ ratio, implying that the local dust
in this position has perhaps been subjected to different environmental 
effects, such as additional photoprocessing.  

We present an analysis of the three dust components along the two sightlines. 
Section \ref{sect:obs_dr} details the observations and data reduction. 
Section \ref{sect:results} summarizes the main observing results and describes a
semi-empirical modelling of the emission across the broad ISOPHOT wavelength range.
FIR opacities and gas column densities are also estimated. 
In Sect. \ref{sect:RTmodels} we present a more detailed physical modelling of the mid- and far-IR 
emission based on radiative transfer calculations.
The discussion of the results in Sect. \ref{sect:Discussion} contains two main issues, (i) the radiation field 
in LDN~1688 and its role for the IR emission, and (ii) the properties of the dust components, 
especially as compared to the diffuse/translucent sightlines in G~300.2~-16.8 (Paper 1). On the basis of
several basic assumptions about geometry and extinction, Appendix \ref{sect:ISRF_model} presents
a plausible description of the local ISRF components.

   \begin{figure}
   \center
   \hspace{-5mm}
      \includegraphics[angle=0,width=8.8cm]{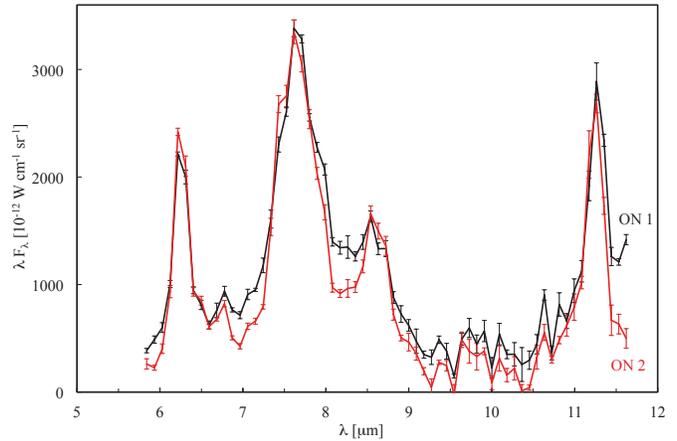}
      \caption{Flux-calibrated, sky-subtracted, averaged PHT-SL spectra for the ON1 (black line) and ON2 (red line) positions.             }
         \label{fig: black_red_specs}
   \end{figure}

   \begin{figure}
   \hspace{-5mm}
   \includegraphics[angle=0,width=9.1cm]{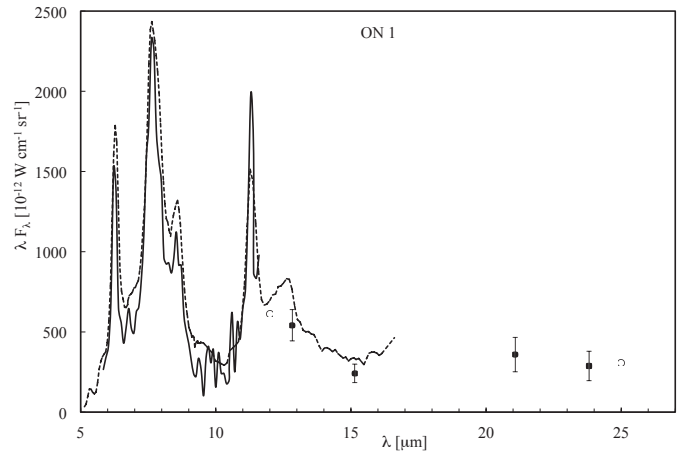}
   \vspace{-3mm}
      \caption{
A comparison of the ISOCAM-CVF spectrum with the ISOPHOT spectroscopy and photometry of the ON1 position (filled circles). The fine solid line is the PHT-S spectrum (with a correction applied to account for differences in pixel binning), the dashed line is the ISOCAM-CVF spectrum. The two IRAS fluxes are also included for comparison (open circles). ISOPHOT data at $\lambda \ge 20 \mu$m have been colour corrected.
              }
         \label{fig: cam_comp}
   \end{figure}

\section{Observations and data reduction}
\label{sect:obs_dr}

\subsection{Spectroscopic observations}
2$\times$2 raster spectra were obtained using PHT-S spectrophotometry mode (AOT 40) \citep{klaas94}.
The TDTs for these observations were 63901741 (OFF), 63901742 (ON2) and 63901743 (ON1). 
The data reduction was performed using the ISOPHOT Interactive Analysis Program (PIA) Version V9.0 \citep{gabriel}.
The OFF position spectrum only exhibited Zodiacal Light (ZL) emission, and no PAH features were seen.
The data were reduced to the AAP level, and then had the OFF position results subtracted.
Single PHT-SL spectra were obtained by using PIA to perform a statistically-weighted
average over the four positions. The final errors were generated automatically by PIA via the
propagation of the initial observational errors. It was found that significant deviation from the
averages of both the continuum and peak levels was seen at the OFF (1,1) raster position. This
was attributed to detector memory effects. The data at this single raster position were therefore
rejected, and the final averaged spectra were instead generated using the three remaining raster
positions. The resultant spectra are shown in Fig.~\ref{fig: black_red_specs}.\\

To obtain an independent check on the overall properties and calibration of the ISOPHOT dataset, 
a comparison with ISOCAM-CVF spectroscopy data obtained covering the ON1 position was conducted. The reduced
and recalibrated ISOCAM datacube (\citealt{boulanger05}; Boulanger, private communication) was
analysed using the Starlink {\sc Gaia} software.

\begin{landscape}
\begin{table}

\caption{Observational ISOPHOT photometry data and accompanying error estimates.}
      \label{tbl: table1}
      \center
\begin{tabular}{lllllllllllll}

\hline
    Filter & $\lambda_{C}$ & $\Delta\lambda$ & Aperture & TDT & $\Delta S_\nu$ & $\sigma_{\rm{INT}}$ &  $\sigma_{\rm{EXT}}$/$\Delta S_\nu$ & Absolute & Filter-To-Filter & Adopted Combined \\
 & & & & On-source (Off-source) & & & & Accuracy & Relative Accuracy & Accuracy \\
 & [$\mu$m]          & [$\mu$m]         &  & & [MJy sr$^{-1}]$ & [MJy sr$^{-1}]$ &[per cent] & [per cent] & [per cent] & [per cent] \\
 (1) & (2) & (3) & (4) & (5) & (6) & (7) & (8) & (9) & (10) & (11) \\
\hline
\multicolumn{11}{|l|}{{\bf ON1:} 16$^{h}$ 25$^{m}$ 40.95$^{s}$ -24$^{\circ}$ 06' 47.0"; (J2000.0)} \\
 & & & & & & & & & & & \\
   P1\_12.8 &      12.83 &       2.33 & 52'' diam. & 83600152 (83600150)  &      23.16$^{\mathrm{a}}$ &        0.98$^{\mathrm{a}}$ & 8 &        20 &         10 & 18 \\
     P1\_16 &      15.14 &       2.86 & 52'' diam. & 83600155 (83600153)  &      12.09$^{\mathrm{a}}$ &        1.93$^{\mathrm{a}}$ & 14 &        20 &         10 & 24 \\
     P2\_20 &      21.08 &       9.43 & 52'' diam. & 83300258 (83300256)  &      25.23$^{\mathrm{a}}$ &       1.46$^{\mathrm{a}}$ & 20 &        20 &         10 & 30 \\
     P2\_25 &      23.81 &       9.18 & 52'' diam. & 83300261 (83300259)  &      22.86$^{\mathrm{a}}$ &       2.32$^{\mathrm{a}}$ & 22 &        20 &         10 & 32 \\
     C1\_60 &       60.8 &       23.9 & 135''$\times$135''  & 83300267 (83300265)  &      169.66 &        1.04 &        8 &        25 &         20 & 33 \\
     C1\_70 &       80.1 &       49.5 & 135''$\times$135''  & 84000170 (84000168)  &      261.93 &        42.92 &         5 &        25 &         20 & 30 \\
    C1\_100 &      103.5 &       43.6 & 135''$\times$135''  & 84000173 (84000171)  &      484.99 &       3.63 &          2 &        25 &         20 & 27 \\
    C2\_135 &        150 &       82.5 & 181''$\times$181''  & 84000176 (84000174)  &      463.28 &       3.16 &         2 &        20 &         10 & 22 (12)$^{\mathrm{b}}$ \\
    C2\_200 &      204.6 &       67.3 & 181''$\times$181''  & 84000179 (84000177)  &      359.95 &       5.03 &         2 &        20 &        10 & 22 (12)$^{\mathrm{b}}$ \\
 & & & & & & & & & & \\
\multicolumn{11}{|l|}{{\bf ON2:} 16$^{h}$ 24$^{m}$ 55.52$^{s}$ -23$^{\circ}$ 00' 50.3"; (J2000.0)} \\
 & & & & & & & & & & \\
     P1\_16 &      15.14 &       2.86 & 52'' diam. & 83600154 (83600153)  &      6.89$^{\mathrm{a}}$ &        1.89$^{\mathrm{a}}$ &     14 &        20 &         10 & 24 \\
     P2\_20 &      21.08 &       9.43 & 52'' diam. & 83300257 (83300256)  &      18.61$^{\mathrm{a}}$ &       1.52$^{\mathrm{a}}$ &    20 &        20 &         10 & 30 \\
     P2\_25 &      23.81 &       9.18 & 52'' diam. & 83300260 (83300259)  &      13.21$^{\mathrm{a}}$ &       2.20$^{\mathrm{a}}$ &    22 &        20 &         10 & 32 \\
     C1\_60 &       60.8 &       23.9 & 135''$\times$135''  & 83300266 (83300265)  &      217.43 &        1.06 &         8 &        25 &         20 & 33 \\
     C1\_70 &       80.1 &       49.5 & 135''$\times$135''  & 84000169 (84000168)  &      322.84 &        50.90 &          5 &        25 &         20 & 30 \\
    C1\_100 &      103.5 &       43.6 & 135''$\times$135''  & 84000172 (84000171)  &      440.40 &       3.65 &        2 &        25 &         20 & 27 \\
    C2\_135 &        150 &       82.5 & 181''$\times$181''  & 84000175 (84000174)  &      409.28 &       3.08 &         2 &        20 &         10 & 22 (12)$^{\mathrm{b}}$ \\
    C2\_200 &      204.6 &       67.3 & 181''$\times$181''  & 84000178 (84000177)  &      265.56 &       5.19 &        2 &        20 &        10 & 22 (12)$^{\mathrm{b}}$ \\
& & & & & & & & & & \\
\multicolumn{11}{|l|}{{\bf OFF:} 16$^{h}$ 19$^{m}$ 40.24$^{s}$ -23$^{\circ}$ 51' 50.9"; (J2000.0)} \\
\hline
\end{tabular}
\begin{list}{}{}
\item[Column descriptions:]{(1) Filter name; 
(2) Central wavelength, $\lambda_{C}$; 
(3) Filter width, $\Delta\lambda$; 
(4) Aperture size and shape; ``diam.'' denotes the diameter of a circular aperture, `135''$\times$135''' and `181''$\times$181''' denote the size of a square detector pixel; 
(5) ISO TDTs of observations, on-source and off-source; 
(6) (ON $-$ OFF) surface brightness, $\Delta S_\nu$; 
(7) Mean PIA-propagated internal error, $\sigma_{\rm{INT}}$, as described in Sect. 2.4 of Paper 1; 
(8) Mean external error (adopted from Paper 1, Sect. 2.4), $\sigma_{\rm{EXT}}$, as a fraction of $\Delta S_\nu$; 
(9) Absolute accuracy (see Klaas et al. 1994); 
(10) Filter-to-filter relative calibration accuracy (see Paper 1); 
(11) Adopted combined accuracy. For the ZL-calibrated PHT-P data, this is taken to be $\sigma_{\rm{EXT}}$ + Filter-to-filter relative calibration accuracy. For the PHT-C data, in the absence of an independent (ZL) calibration curve, this is taken to be $\sigma_{\rm{EXT}}$ + absolute accuracy.}
\item[$^{\mathrm{a}}$]{Values listed for the P1 and P2 bands are after ZL calibration.}
\item[$^{\mathrm{b}}$]{The filter-to-filter relative calibration accuracy values listed for the C2 data were used to produce the error estimates in parentheses. These were used when estimating uncertainties on FIR temperatures (see Sect. 5.1 of Paper 1). These smaller values are justified for this purpose, as the two sets of measurements all use the same detector configuration and only differ in the filter used, and are thus not susceptible to cross-detector calibration uncertainties. The quoted values are based on an estimate of upper limits on remaining uncertainties due to responsivity variations derived in Paper 1.}
\item[$^{\mathrm{c}}$]{Errors are listed in parentheses.}
\end{list}
\end{table}
\end{landscape}

\subsection{Photometric observations}
The photometric observations were 
performed in the filters that are listed in Table~\ref{tbl: table1}, along with their corresponding PHT-P
aperture diaphragm sizes or the PHT-C camera field sizes and the observation TDTs.
The region corresponding to the ON1 position was sampled
over a square area matching the ISOPHOT aperture in both size and position, enabling an averaged
spectrum to be extracted. The results are shown in Fig.~\ref{fig: cam_comp}.  If an ISO-typical error of 17 per cent on the
absolute flux levels is assumed for both spectroscopy datasets, the spectra are found to be
in agreement, both in the IR band peaks and the continuum regions. The sparse map observing
templates (AOTs 17/18/19: PHT-P and 37/38/39: PHT-C) were used for photometry \citep{klaas94} for each filter.

\subsection{Photometry reduction and error analysis}
The photometric data reduction was performed using
PIA Version V9.0 \citep{gabriel}. The data reduction process as
described in Paper 1 was used: the data were calibrated against the
on-board calibration source FCS1, and the ZL. This latter was achieved via the
ISOPHOT ZL template spectra of \citet{leinert} and the colour-corrected monochromatic
{\it COBE}/DIRBE fluxes at 3.5, 5, 12 and 25 $\mu$m, under the assumption of blackbody behaviour.
Signal drift correction and foreground subtraction were also performed as in Paper 1. In all of
the filter bands, a clear excess signal is seen towards the ON positions. As in this earlier work, we
list here estimate of internal and external errors. Unlike for the G~300.2~-16.8 photometry dataset,
only single ON and OFF photometry measurements were obtained, making a statistical average calculation
of the external errors, $\sigma_{\rm{EXT}}$, not viable.  However, the fractional external systematic
errors are not expected to be larger than those in Paper 1.
Since both these errors and the systematic calibration errors  \citep{klaas02} here arise from the same
causes, we therefore adopt the statistical and systematic error estimates used in Paper 1 for the
corresponding data in this work.

The adopted combined errors for the average (ON -- OFF) signal
listed in column 11 of Table~\ref{tbl: table1} were obtained by arithmetically adding the statistical
external errors (column 8) and the filter-to-filter systematic (calibration) errors (column 10).

When estimating the filter-to-filter accuracies at $\lambda \ge 60 \mu$m and the absolute accuracies at
all wavelengths, the results of the \citet{klaas02} investigation of ISOPHOT accuracies were adopted.
The estimated absolute accuracies are given in column 9 of Table~\ref{tbl: table1}. As in Paper 1, although the
absolute accuracies for $\lambda \ge 60 \mu$m are $\sim$ 20 $-$ 25 per cent, the filter-to-filter
uncertainties for a {\it single detector} (e.g. C2) may be substantially smaller due to the elimination of
most of the sources of error (see \citealt{delBurgo}, Sect. 4). Unlike the shorter-wavelength data, the
absence of any independent calibration curve to aid cross-calibration for the $\lambda \ge 60 \mu$m data
dictated that the adopted combined errors for the average (ON -- OFF) signal listed in column 11 of
Table~\ref{tbl: table1} be obtained by arithmetically adding the statistical external errors (column 8) and the
absolute errors (column 9).

\subsection{Extinction Mapping of LDN 1688}
Although LDN 1688 is expected to be opaque at $V$ and $I$,  the 2MASS $JHK_{S}$ data can
be used to derive near-infrared (NIR) colour excesses of stars visible through LDN 1688, and hence
the NIR extinction. As in \citet{lombardi2008} and Paper 1, we have applied the optimized multi-band
technique of \citet{lombardi}, adopting $A(J) / E(J - H) = 2.66$ and $A(J) / E(H - K_S) = 4.13$,  
following the extinction curve of \citet{mathis90}.
The reference area for setting the absolute extinction levels was a 10'-radius region at
16$^{h}$ 0$^{m}$ 10.0$^{s}$ -24$^{\circ}$ 32' 0.0"; (J2000.0).
The resultant extinction map is shown in the lower-right panel of Fig.~\ref{fig: WISE_Herschel_AJ_ims}. The extinction for each map pixel
was derived from the individual extinction values of stars by applying the sigma-clipping smoothing
technique of \citet{lombardi} and using a Gaussian with FWHM = 4' as a weighting function for the
individual extinctions. This produced extinctions of $A(J) = 1.88 \pm 0.06$ at the ON1 position and
$A(J) = 0.98 \pm 0.06$ at the ON2 position. The measured extinction at the OFF reference position was $A(J) = 0.56 \pm 0.06$.
Subtracting this from the ON position extinction values yielded $A(J) = 1.32 \pm 0.08$ for (ON1 -- OFF) and $A(J) = 0.42 \pm 0.08$ for (ON2 -- OFF).

\section{ISOPHOT results and semi-empirical modelling}
\label{sect:results}

\begin{figure*}
   \includegraphics[angle=0,width=17.9cm]{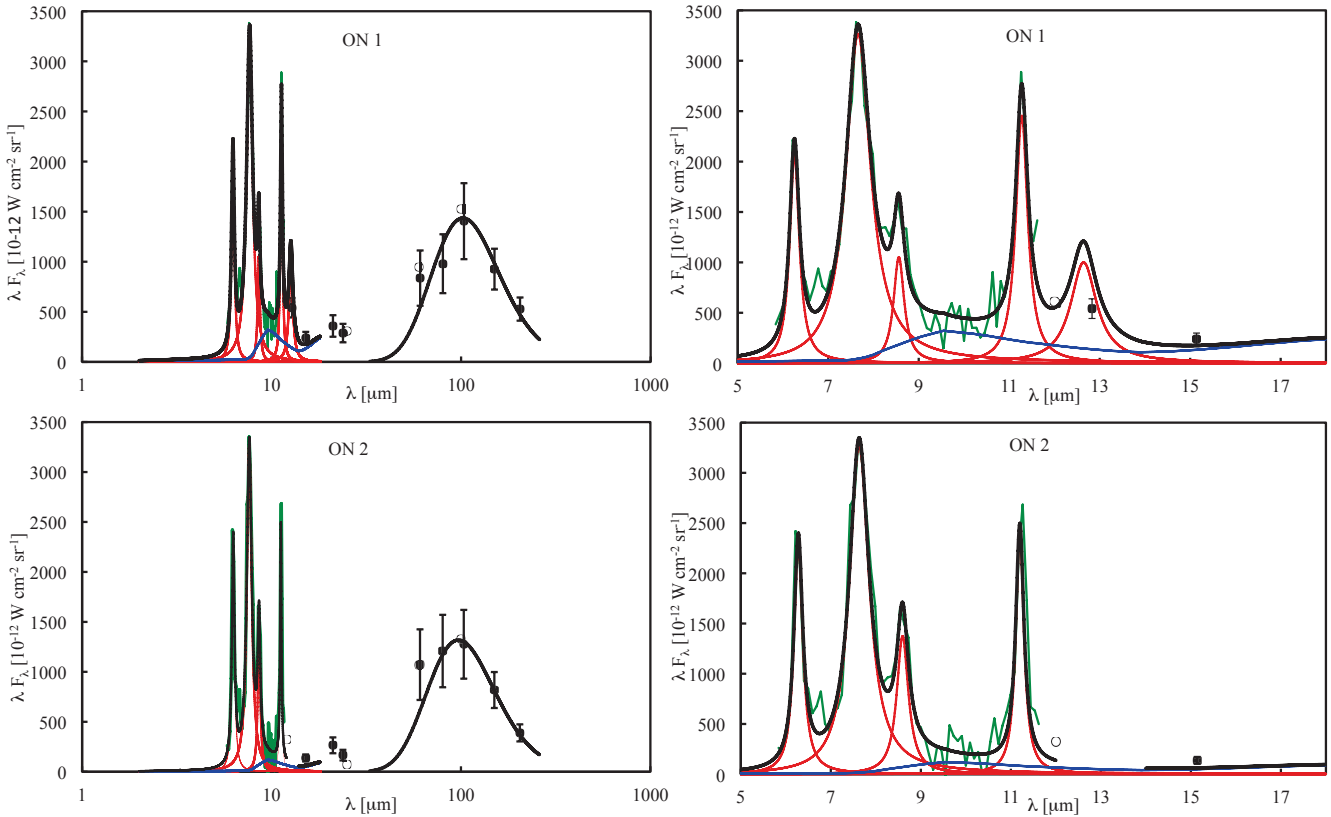}
            \caption{
Semi-empirical fitting results. In the left-hand panels, modified blackbody fits of the form  $\nu ^{2} B_{\nu} (T)$ to the $100-200$ $\mu$m data points is shown as a black line spanning the $40-250$ $\mu$m wavelength range (ON1: upper panels, ON2: lower panels). The PHT-SL spectra are shown in green. Cauchy band profiles fitted to the $6.2-12.7$ $\mu$m features are shown in red. A \citet{li01b}-style silicate continuum is shown in blue for the $2-18$ $\mu$m range, and a combination of these is shown as a black line for $2-18$ $\mu$m. The right-hand panels show the details of the emission feature fitting results.
              }
         \label{fig: semi_emp}
   \end{figure*}

Figs.~\ref{fig: black_red_specs} --~\ref{fig: semi_emp} show the data for the two observed positions as spectral energy distributions. 
The MIR spectra exhibit strong 7.7- and 11.3-$\mu$m features. The underlying continuum emission, longward of $\sim$10 $\mu$m and noted in Paper 1, is also present in LDN~1688.

As detailed in Paper 1, a semi-empirical fit to the MIR spectra is performed
using a combination of Cauchy curves and a silicate continuum to model the
molecular-level IR band emission, together with a modified blackbody fit for the thermal
emission at longer wavelengths by large grains.

For the IR band emission, the fits are $\chi{^2}$ minimization fits to the PHT-SL data wherever available.
For wavelengths longward of the PHT-SL wavelength coverage, the areas under the
fitted curve were constrained to fit the colour-corrected integrated 12.8 (where available) and 16$\mu$m
ISOPHOT filter fluxes. For both sightlines, the resultant fits (shown
in Fig.~\ref{fig: semi_emp}) reproduce the in-band integrated fluxes within 20 per cent of the
photometrically-measured values. 

The three longest-wavelength photometry points, C100, C135 and C200 are fitted
using a modified blackbody function of the form $\nu^{2} B(\nu)$. This function is
individually convolved with each of the three ISOPHOT filter response curves in turn,
and the temperature and scaling are adjusted to best fit the photometry. At positions
ON1 and ON2, the temperatures of $\sim$23.5 K and  $\sim$24.8 K, respectively, were obtained (Table~\ref{tbl: table2}).

The resulting fits are shown in Fig.~\ref{fig: semi_emp} and the parameters are listed in Table~\ref{tbl: table2}. It can
be seen from Fig.~\ref{fig: semi_emp} that in both cases, the middle range of the spectrum
($\sim$25 $-$ 70 $\mu$m) exhibits additional emission that can be explained by
 the presence of a transiently-heated grain component.
Physical grain population modelling is therefore required, and is described in Sect. 4. We also note
that the continuum near 10 $\mu$m appears to be a much less significant contribution
than in Paper 1.

\begin{table} 
\center
   \caption[3]{Results of semi-empirical model fitting. Overall
errors on the peak heights are estimated to be $\sim$ 17 per cent based on a comparison of the ON1 spectrum with the 
ISOCAM CVF spectrum, with relative band-to-band
errors $\sim 10-15$ per cent (as in Paper 1). $F(\lambda )$ 
values are quoted in units of $10^{-12}$W cm$^{-2} \mu \rm{m}^{-1} \rm{sr}^{-1}$.}
      \label{tbl: table2}
\begin{tabular}{llll}
\hline 
Feature & Fit parameter &        ON1 &        ON2 \\
\hline
6.2-$\mu\rm{m}$ & Central $\lambda$ $[\mu\rm{m}]$ &       6.25 &       6.28 \\

band & Width$[\mu\rm{m}]$ &       0.26 &      0.24 \\

           & $F(\lambda)$ Height &    329.5 &     364.4 \\
\\
7.7--$\mu\rm{m}$ & Central $\lambda$ $[\mu\rm{m}]$ &       7.65 &       7.62 \\

band & Width$[\mu\rm{m}]$ &       0.66 &      0.53 \\

           & $F(\lambda)$ Height &    426.6 &     430.0 \\

\\
8.6--$\mu\rm{m}$ & Central $\lambda$ $[\mu\rm{m}]$ &       8.56 &       8.59 \\

band & Width$[\mu\rm{m}]$ &       0.28 &      0.31 \\

           & $F(\lambda)$ Height &    122.9 &     160.0 \\
\\
11.3--$\mu\rm{m}$ & Central $\lambda$ $[\mu\rm{m}]$ &       11.28 &       11.20 \\

band & Width$[\mu\rm{m}]$ &       0.32 &      0.22 \\

           & $F(\lambda)$ Height &    217.7 &     212.5 \\
\\

12.7--$\mu\rm{m}$ & Central $\lambda$ $[\mu\rm{m}]$ &       12.64 &       -- \\

band & Width$[\mu\rm{m}]$ &       0.7 &      -- \\

           & $F(\lambda)$ Height &    79.1 &     -- \\
\hline
\multicolumn{ 2}{l}{7.7/11.3-$\mu\rm{m}$ band ratio} & $2.0\pm0.3$ & $2.0\pm0.3$ \\
\hline
\multicolumn{ 2}{l}{10-$\mu\rm{m}$ continuum level} & $29.9 \pm 2.2$ & $11.2 \pm 0.8$ \\
\multicolumn{ 3}{l}{(average of $F(\lambda)$ over $9.5-10.5$ $\mu\rm{m}$ range)} & \\
\hline
\multicolumn{ 2}{l}{Classical grains: equilibrium} &       23.5 &       24.8 \\
\multicolumn{ 2}{l}{temperature, $T$ [K]} & $\pm^{2.1}_{1.9}$ & $\pm^{2.2}_{2.0}$ \\
\hline
\end{tabular}
\end{table}

\begin{table}
\center
   \caption[5]{Visual extinction, optical depth and average absorption cross-sections
per H-nucleon for ON1 and ON2. $\tau_{200}$ is calculated under the assumption
that $\tau_{\lambda} \propto \lambda^{-2}$.}
      \label{tbl: table3}
\begin{tabular}{lll}
\hline
           &        ON1 &        ON2 \\
\hline
   $A(J)$ & $1.32 \pm 0.08$ & $0.42 \pm 0.08$ \\

$N(H)$ $[10^{21}$ cm$^{-2}]$ $$ &       $7.4 \pm 0.7$ &        $ 2.4 \pm 0.5$ \\

$I_{200} (\nu)$ [MJy sr$^{-1}$] &       $360.0 \pm 79$ &       $266.6 \pm 59$ \\

$\tau_{200}$ $[10^{-4}]$ & $14.8 \pm ^{5.5}_{4.9}$ & $9.2 \pm ^{3.3}_{3.0}$ \\

$I_{200} / A(J)$ [MJy sr$^{-1}$] &        $273 \pm 62$ &       $635 \pm 62$ \\

$\tau_{200} / A(J)$ $[10^{-4}\rm{ mag}^{-1}]$ &   $11.2 \pm ^{4.2}_{3.7}$ &   $22.0 \pm ^{9.0}_{8.2}$ \\

$\sigma_{200}^{\rm{H}} = \tau_{200} / N(H)$ & $2.0 \pm 0.8$ & $3.9 \pm 1.6$ \\
$[10^{-25}\rm{~cm}^{2}\rm{~H~nucleon}^{-1}]$ & & \\
\hline
\end{tabular}
\end{table}

\subsection{Far-Infrared Opacity}
\label{subsect:FIRopacity}
 
We derive here an estimate of the ratio between the FIR optical depth 
$\tau_{\rm{em}}(\lambda)$ at 200 $\mu$m and the J-band optical extinction $A(J)$, 
$\tau_{200}/A(J)$, as well as the average absorption cross section per H-nucleon, 
$\sigma_{\lambda}^{\rm{H}} = \tau(\lambda) / N(H)$.

For the case of optically thin emission and an isothermal
cloud, the observed surface brightness is $I(\lambda) = \tau(\lambda) B(\lambda, T_{\rm{dust}})$.
Using the dust temperature values $T_{\rm{dust}}$ for ON1 and ON2  as given in Table~\ref{tbl: table2},
we have hence calculated the optical depths.

The total hydrogen column density $N(H)$ is obtained from the
$A(J)$ values using the the $N(H)$ vs. $A(J)$ relationship as determined 
by \citet{vuong2003}
from X-ray absorption measurements of background stars in LDN~1688,  
$N(H)/A(J) = 5.57\pm 0.35 \times 10^{21} \rm{cm}^{-2} \rm{mag}^{-1}$.
The resulting N(H) and $\sigma_{200}^{\rm{H}}$ values are given in Table~\ref{tbl: table3}.
\citet{vuong2003} noted that the ratio for the  the general Galactic ISM is
$N(H)/A(J) = 6.4 - 7.8 \times 10^{21} \rm{cm}^{-2} \rm{mag}^{-1}$,
i.e. some 15 to 40 per cent  higher than for LDN~1688 (a $2\sigma$ effect).

\section{Physical modelling}
\label{sect:RTmodels}

As an alternative to the semi-empirical fits presented above, we construct here
radiative transfer models to explain the observations of the positions ON1
and ON2. We use spherically symmetric models that are intended to describe
the average properties of these regions. The radiation field consists of
the standard ISRF \citep{mathis83} that is scaled,
as necessary, to represent the higher intensities found in the cloud. We
also examine the effect of an additional radiation field component that
corresponds to the spectrum of a B3V star \citep{bruzual2003}. The
modelling is completed using the dust model of Li \& Draine (2001),
hereafter referred to as the LD model. The LD model consists of PAHs,
graphite grains, and silicate grains. 

The radiative transfer models were optimised to reproduce the
observations of dust emission and extinction. In the $\chi^2$
minimisation, the photometric measurements were weighted according to the
error estimates. Below 12\,$\mu$m, we use the PHT-SL spectrum. Its weight
is divided by the number of the frequency points to reduce its overall
influence on the fits. We require that the opacity of the models (as
measured through the centre of the model clouds) corresponds to
observations. The weight of this constraint was determined by assuming a
20 per cent relative uncertainty for the observed $A(J)$ values.

\begin{figure}
\includegraphics[width=8cm]{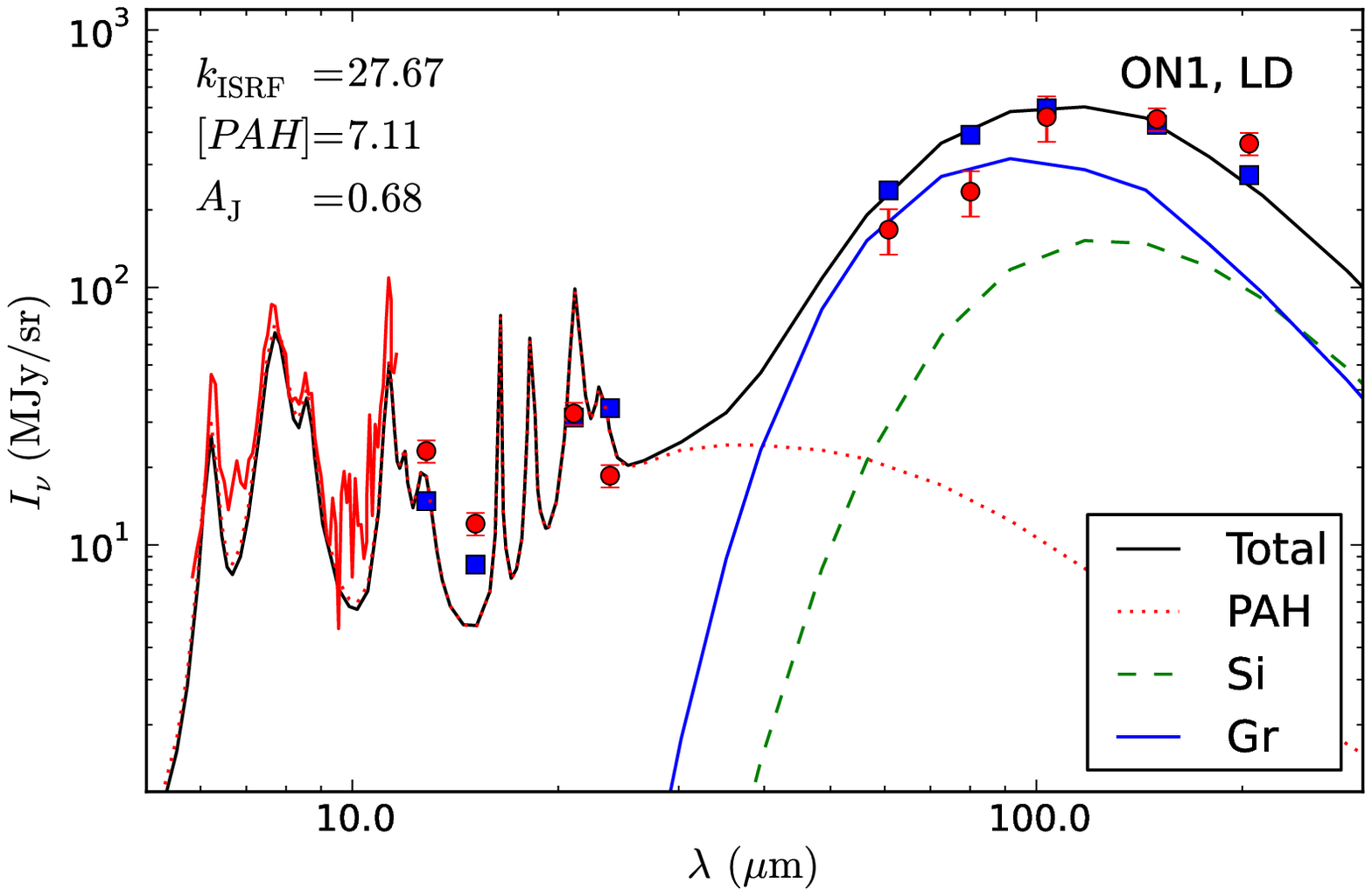}
\includegraphics[width=8cm]{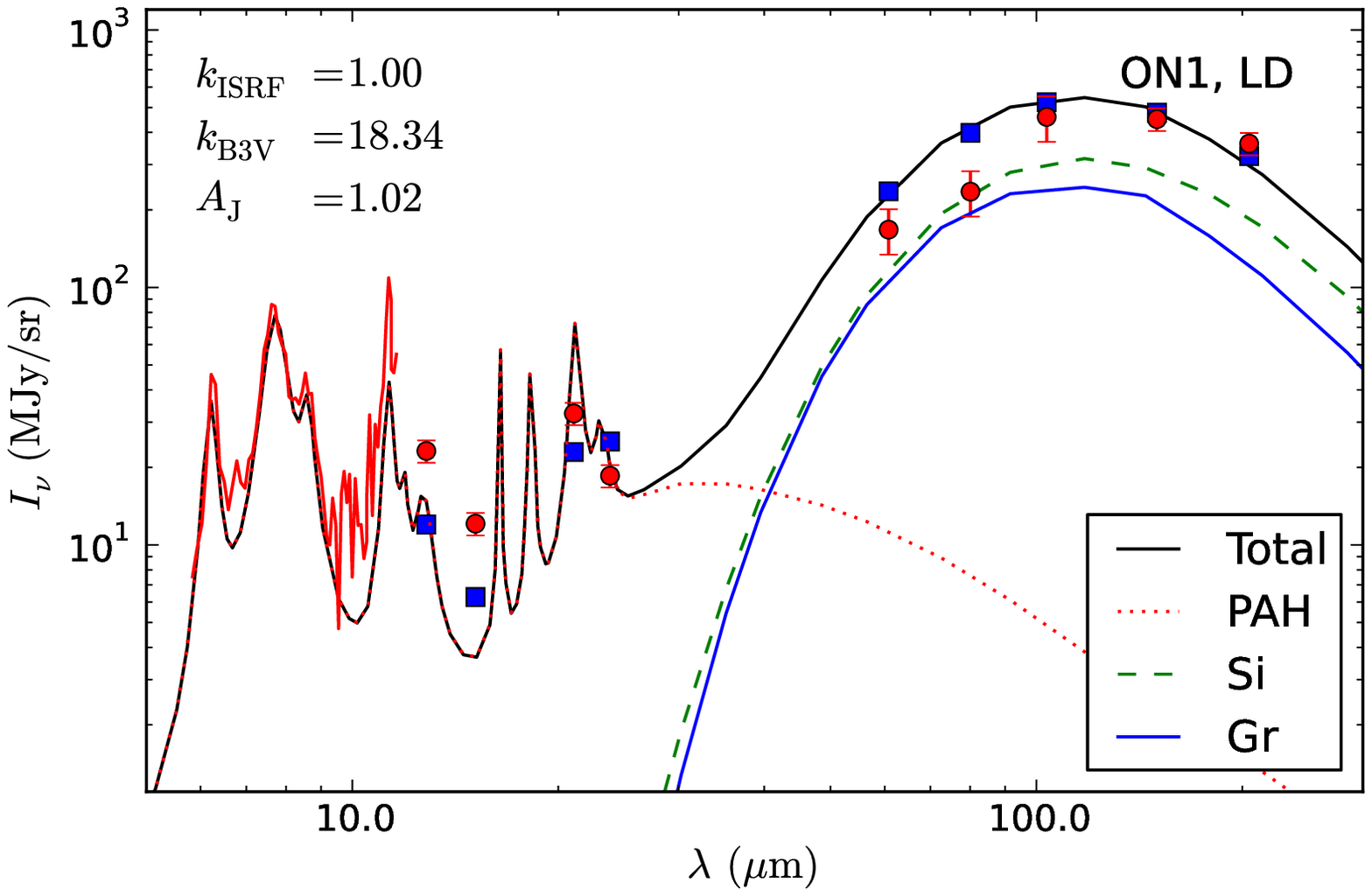}
\caption{
Model fits for position ON1 using the Li \& Draine (2001) dust model.  The
lines show the three emission components and the total intensity calculated
from the model. The red symbols and the red line show the photometric
observations (without colour correction, i.e., values at the reference
wavelength assuming a spectrum $\nu I_\nu =$ constant) and the observed
PHT-SL spectrum. The blue symbols are the corresponding values that are
calculated from the model spectrum for the various ISOPHOT filters. The
free parameters include $k_{\rm ISRF}$ and $k_{\rm PAH}$ in the upper frame
and $k_{B3}$ in the lower frame. The total column density is another free
parameter but restricted by the observed value of $A(J) = 1.2$.
}
\label{fig:ON1_LD}
\end{figure}

\begin{figure}
\includegraphics[width=8cm]{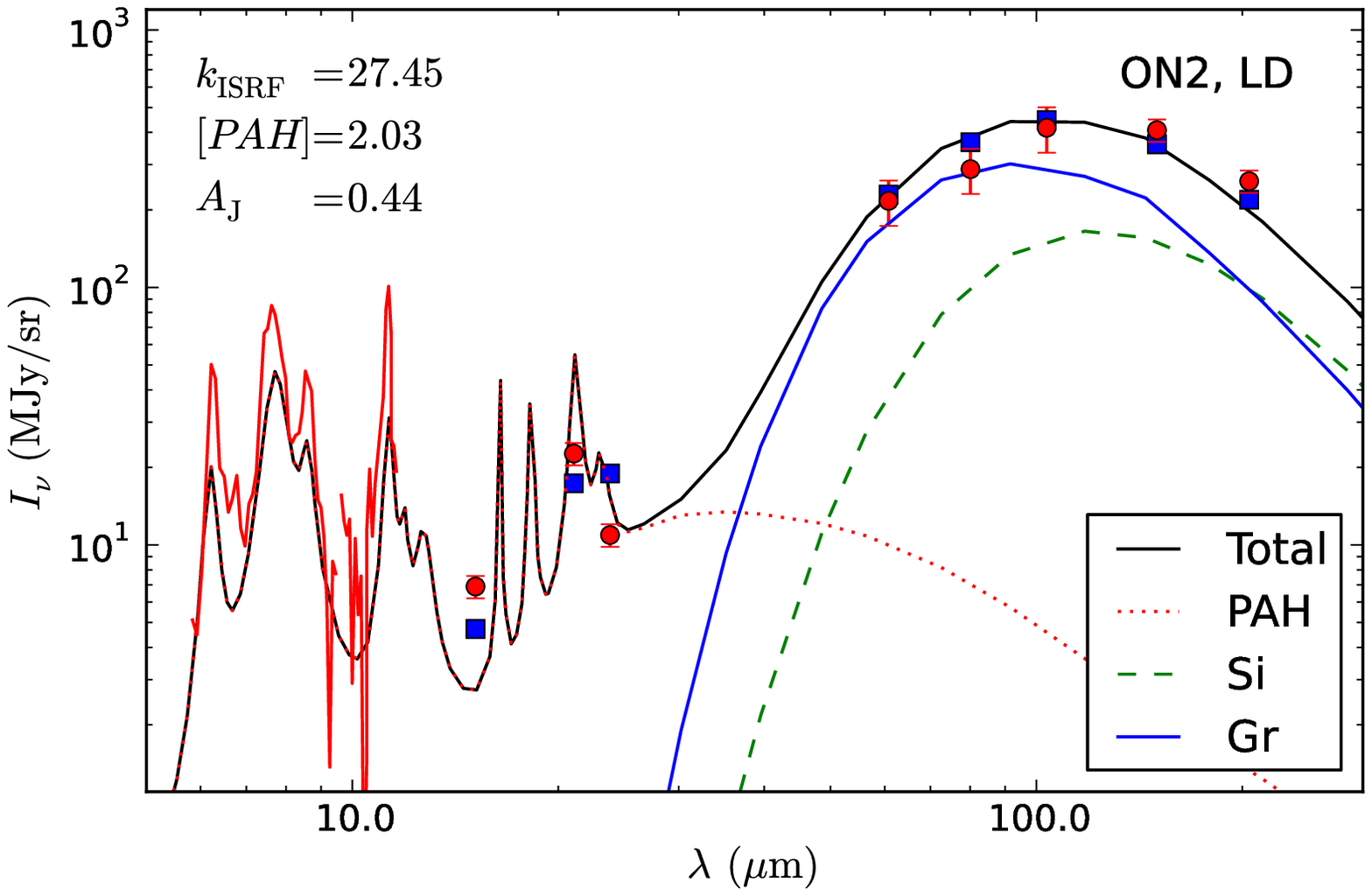}
\includegraphics[width=8cm]{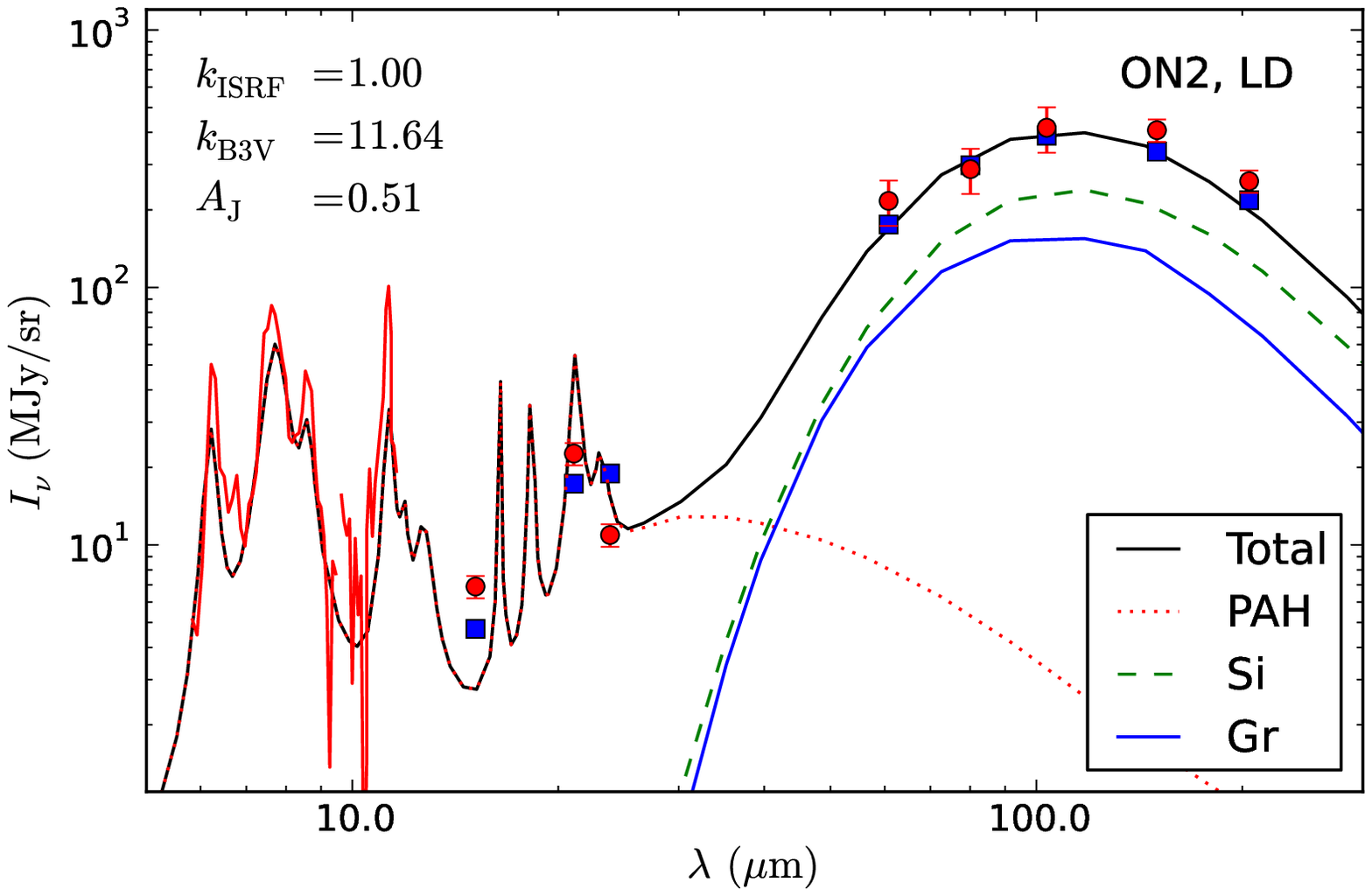}
\caption{
Model fits for position ON2 using the Li \& Draine (2001) dust model
(see Fig.~\ref{fig:ON1_LD} for details). The target value for 
$A(J)$ is 0.39.
}
\label{fig:ON2_LD}
\end{figure}

The LD models were constructed in part similarly to Paper 1.  The model
column density was kept as a free parameter although restricted by the
$A(J)$ measurements.
In the first models, the scaling of the ISRF, $k_{\rm ISRF}$, and the
relative abundance of the PAHs are treated as additional free parameters
(upper frames in Figs.~\ref{fig:ON1_LD} and \ref{fig:ON2_LD}). In the
best fits, the PAH abundance and the radiation field intensity are raised
far above the default values. For ON1, the best fit is obtained with
$k_{\rm ISRF}=30$ and with the PAH abundance increased by a factor of
$\sim$5. For the ON2 position, the corresponding numbers are $\sim$27 and
$\sim$2. The opacity of the model clouds, as measured by $A(J)$,
is less than half of the target value  for ON1, while for ON2, $A(J)$
is $\sim$10 per cent above the value derived from observations.

\begin{figure}
\includegraphics[width=8.3cm]{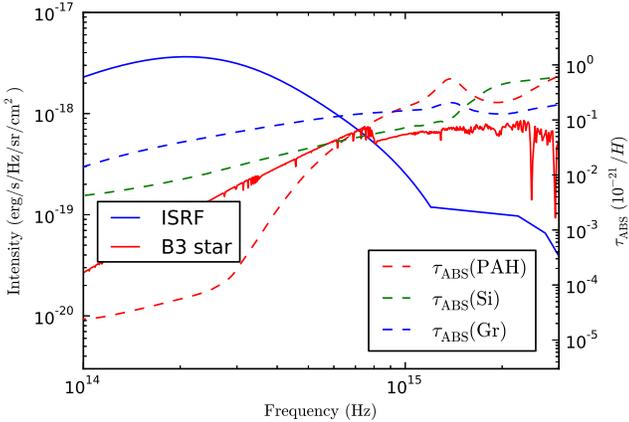}
\caption{
SEDs of the local ISRF and B3V stellar emission, together with the
excitation cross-sections for the Li \& Draine (2001) model components. The
``PAH" line represents the PAHs and the small graphite grains (below 50
\AA~in size).
}
\label{fig:ISRF_LD}
\end{figure}

In the second set of models, the ISRF was fixed to the Mathis et al. (1983)
value but we added another radiation field component that corresponds to
the spectrum of a star with a spectral class of B3V. 
Fig.~\ref{fig:ISRF_LD} compares the SED of the local ISRF with the spectrum
of a B3V star. The same figure shows the extinction cross-sections for the
dust components of the LD models. B3V star is clearly an efficient source
of dust heating, especially in the case of the PAHs.
In the modelling, the intensity of the B3V component was scaled with a
factor $k_{\rm B3V}$, where a value of 1.0 corresponds to a radiation
field energy equal to that integrated over the full ISRF spectrum
\citep{mathis83}. The free parameters are thus the column density of the
model and the factor $k_{\rm B3V}$ (as indicated in the lower frames of
Figs.~\ref{fig:ON1_LD} and \ref{fig:ON2_LD}). For ON1, the $\chi^2$
minimum is reached with $k_{\rm B3V}$ = 18, while for ON2, the value is
$k_{\rm B3V}$ = 12. For the position ON1, the $A(J)$ value is now
within 20 per cent of the observed value, while for ON2 the value is $\sim$30 per cent
above the observed value. The difference in the quality of the fits
between the first and second set of models is not very significant.
Nevertheless, the $\chi^2$ values are lower by $\sim$25 per cent for both ON1
and ON2.

We also tested the case in which both $k_{\rm ISRF}$ and $k_{\rm B3V}$ were
kept as free parameters. The results are very similar to the case in which
only $k_{\rm B3V}$ was varied but the value of $k_{\rm ISRF}$ decreases
below one. However, the $k_{\rm ISRF}$ values are no longer well
constrained because the radiation field is completely dominated by the
B3V component.

\section{Discussion}
\label{sect:Discussion}

\subsection{The Interstellar Radiation Field in LDN~1688}
\label{sect:ISRF_model}

Our modelling in Section 4 of the 6 - 200 $\mu$m emission at the two sight
lines ON1 and ON2 (Figs. \ref{fig:ON1_LD} and \ref{fig:ON2_LD}) indicated that,
in order to fit the far-IR region, a radiation field enhanced by a factor of
$\sim$ 20 - 30 is needed if a normal Solar Neighbourhood ISRF SED shape
\citep{mathis83} is assumed. The LDN~1688 cloud is, however, known to contain
embedded early B type stars with an SED strongly increasing towards UV.

There are two other sources contributing to the ISRF  in LDN~1688:

(i) near to mid-IR radiation from the rich embedded cluster of newly formed stars in the central
molecular and dust core \citep{wilking2008}, and

(ii) external UV-optical illumination from bright early type stars of the Upper Sco OB association,
mainly from the northern side of the Ophiuchus cloud complex.

Estimates of the three ISRF components are presented in Appendix \ref{sect:ISRF_model2}.

Among the three  local components the dominating one is the radiation of the two
embedded stars $\rho$ Oph AB and HD~147889. The local ISRF also dominates
over the general ISRF \citep{mathis83} by one order of magnitude.
It is difficult to get an estimate for the intervening {\it effective}
extinction in a scattering and clumpy medium between star and the line-of-sight
dust. Therefore, we are not able to get a quantitative estimate better than
within a factor of 3 -- 4 of the modelled ISRF. Within these uncertainties, however, the
ISRF model and the observationally derived values (Sect.~\ref{sect:RTmodels}) are in
qualitative agreement with each other.

\subsection{The dust properties}

The ISOPHOT dataset for LDN~1688 describes the IR emission spectrum over a
large wavelength range, encompassing emission from at least three dust
grain populations, as also seen in Paper 1 in the case of the cloud
G~300.2~-16.8. The column densities on the LDN~1688 lines-of-sight are
higher, but by no more than a factor of 2 -- 3. The radiation field
is significantly higher in LDN~1688, however, as indicated by the FIR emission that is
one order of magnitude higher than that of the the cloud G~300.2~-16.8 (see also
\citealt{boulanger96a}). The different environments could be reflected in
the properties of the emission from the (possibly ionized) aromatic
molecular species, transiently-heated VSGs and big grains.


\subsubsection{Mid-infrared emission features.}
The MIR spectra at the positions ON1 and ON2 (see Fig.~\ref{fig: black_red_specs}) show
some differences but no large deviations are observed even compared to
other diffuse and translucent sightlines (the cloud G~300.2~-16.8
included). The locations and widths of the PAH features are typical of
those found in the ISM, including regions illuminated by nearby stars
(see \citealt{tielens2008}). A comparison of the semi-empirical fitting
results, listed for LDN~1688 in Table~\ref{tbl: table2}, and the diffuse
sightlines examined by \cite{kahanpaa2003} also show good agreement
between the central wavelengths and widths of the emission features
with the only exception that the width of the 7.7-$\mu$m feature is
smaller in LDN~1688. The 12.7-$\mu$m peaks also appear to be narrow in
LDN~1688 but the feature is outside the ISOPHOT spectroscopic range,
and is also the most likely feature to be affected by uncertainties
associated with the presence of underlying small grain continuum
emission. 

In LDN~1688, the 7.7 / 11.3 $\mu$m band ratio (ratio of the peaks of the
features) is $\sim 2.0$ in both positions. This is lower that the values
2.6 -- 3.4 found in G~300.2~-16.8. The result can be affected by details in
the fit of the underlying continuum which is very strong in the case of
G~300.2~-16.8. In G~300.2~-16.8, the continuum estimates depend mainly on
the 10\,$\mu$m photometric measurements whose relative uncertainty was
$\sim$25 per cent. The differences between LDN~1688 and G~300.2~-16.8 7.7 /
11.3 $\mu$m ratios therefore remain significant. It is well within the
variation seen on other Galactic ISM lines-of-sight \citep{chan2001,
tielens2008}, however, and is small compared to the variation between galaxies
\citep{galliano2008}. The ratios can be attributed to differences in the
ionization / recombination rates of the carrier species arising from the
differences in ambient radiation field strengths.


\subsubsection{Mid-infrared continuum.}
As in Paper 1, our semi-empirical modelling includes a small grain
component to accommodate the presence of a continuum around 10 $\mu$m. This
appears to be correlated with the mid-IR emission at 16 - 25 $\mu$m of the
aromatic emission features. The region around 10\,$\mu$m is traced more
reliably thanks to the availability of PHT-SL spectra and the additional
corroboration from the ISOCAM-CVF measurement at the position ON1
(Boulanger et al. 1996a).  The continuum level is higher at position ON1
than at ON2. This is most clearly visible in the 10\,$\mu$m region, but can also
be recognised as higher values around 7.0\,$\mu$m and 8.1\,$\mu$m between
the PAH features. In contrast, the total signals (continuum + feature) at
the peak positions of the PAH features (6.2, 7.7, 8.6, and 11.3\,$\mu$m)
are practically the same for both ON1 and ON2.

\cite{kahanpaa2003} studied a number of Galactic diffuse
lines-of-sight and, using the average spectrum, derived an upper limit
of 10 per cent for the 10$\mu$m continuum relative to the peak flux density
of the 7.7\,$\mu$m feature. Using the average signal between 9.5 and
10.5\,$\mu$m as the continuum value for comparison, the ratio in LDN~1688 is 9 per cent for
ON1 and 5 per cent for ON2. There is therefore no indication that the
LDN~1688 sightlines significantly differ from the general ISM in this respect.
In terms of the column density, G~300.2~-16.8 falls between LDN~1688
and the case of diffuse sightlines. There, however, the ratios of the
10\,$\mu$m continuum and the 7.7\,$\mu$m peak flux density were
$\sim$20 per cent or above. 

The stronger continuum at the ON1 positions of both LDN~1688 and (especially)
G~300.2~-16.8 must be produced by VSGs or possibly even PAH-sized
particles. The increased emission could result either from a higher
abundance of such particles or from a greater radiation field
intensity. In LDN~1688, the strong radiation of early B-type stars
combined with the high extinction at the ON1 position could
conceivably change the relative intensity of the MIR continuum and the
UV-excited PAH features. A strong effect cannot be expected, however,
because both emission components cover the same wavelength range. In
G~300.2~-16.8, the radiation field is normal, and any explanation of
the continuum must involve changes in the dust properties.

\subsubsection{Comparison of MIR and FIR emission}


The $IRAS$ 12\,$\mu$m / 100\,$\mu$m and (especially) the 25\,$\mu$m/100\,$\mu$m
surface brightness ratios had suggested strong differences between the LDN~1688
sightlines, with the ON2 position showing lower MIR emission. \citet{boulanger96a}
already estimated that at position ON1, $\sim$60 per cent of the $IRAS$
12\,$\mu$m signal could be attributed to the continuum. The $IRAS$ data are
generally consistent with the higher spatial resolution ISO data. However,
the 25\,$\mu$m $IRAS$ measurement of the ON2 position is low compared to ISOPHOT
P2\_20 and P2\_25 filter measurements and, therefore, the difference
between ON1 and ON2 is smaller than in $IRAS$ data. The difference in the
spatial resolution should not play a significant role. The higher
resolution WISE and Herschel maps show that the ON2 position coincides with
a peak in the surface brightness (see Fig.~\ref{fig: WISE_Herschel_AJ_ims}) but
the values vary by less than 20 per cent within the larger area covered by the IRAS beam.

We can compare the LDN~1688 and G~300.2~-16.8 sightlines
quantitatively using the ISOPHOT filters C\_100, C\_135, and C\_200 to
trace the FIR emission and the filters P\_16, P\_20, and P\_25 to trace the
MIR emission longward of the main PAH features. With the in-band power
values (see Table 2 in Paper I), we obtain MIR/FIR ratios of $0.39 \pm 0.10$,
$0.44 \pm 0.09$, and $0.15 \pm 0.04$ for the three positions in
G~300.2~-16.8. In LDN~1688, the corresponding ratios are $0.25 \pm 0.06$
and $0.19 \pm 0.05$ for positions ON1 and ON2, respectively. Thus, the
LDN~1688 sightlines are very similar to the position ON3 in
G~300.2~-16.8 regarding the ratio of MIR and FIR intensity, in spite of
the very different radiation field. In LDN~1688, the MIR / FIR ratio is lower
in position ON2 than in position ON1, the difference being only at a
$1\sigma$ level. 
Using the FIR part of the spectra, we derived a large grain temperature of
$\sim$24 K in both LDN~1688 positions. Based on the SED shape alone, there
would not seem to be a large difference in the intensity of the heating
radiation field. However, one must also take into account the difference in
the opacities. For position ON1, the total optical depth was estimated
to be twice as high as for the ON2 sightline. This would allow for a larger
amount of cooler dust that is shielded from the external UV field.


\subsubsection{Dust properties and the radiation field.}
The physical modelling of Sect.~\ref{sect:RTmodels} helps us to address the
question of the relative importance of the intensity and spectrum of the
heating radiation field and the possible variations in dust properties. By
taking into account dust temperature variations within the model cloud, it
could also result in slightly different estimates of $\tau_{200}$.
We examined the heating by B stars relative to the illumination by a
similarly energetic radiation field with an SED corresponding to the normal
ISRF. The spectra of both LDN~1688 sightlines could be fitted
satisfactorily by assuming a normal ISRF-type SED scaled by a factor
$\sim 30$, and by increasing the relative abundance of PAH grains. For
position ON1, the PAH abundance had to be increased almost by a factor of
$\sim 7$, whereas for position ON2, a factor of $\sim 2$ was sufficient. 
As discussed above, however, the radiation field is known to be affected by the high
mass stars inside the $\rho$ Oph cloud and by the Upper Scorpius OB
association. This justified the alternative models in which the radiation
field contained a significant harder component corresponding to a B3V star.
With this additional radiation field component, not only could the
observations be fitted equally well, but in addition, no modification was needed in
the dust model. This demonstrates the general degeneracy between the
properties of the radiation field and the dust grain abundance. In the case
of LDN~1688, it also suggests that the differences in the environmental
conditions (the radiation field in particular) are not necessarily strongly
reflected in the relative abundances of large grains, VSGs and PAHs.
Nevertheless, one can note that the models underestimate the intensity
between 10--20\,$\mu$m, i.e., a stronger continuum component is needed.
Furthermore, while the models with a B3V stellar component and normal PAH
abundance could fit the shortest wavelengths very well, the observed
intensities at ON1 are underestimated towards the long wavelength end of
the PHT-SL spectrum. The effect is rather subtle but may be another
indication of the need for a larger VSG contribution. It could also suggest
that the employed radiation field is already too hard, making the continuum
too flat below 10\,$\mu$m. The actual LDN~1688 spectra may thus require not
only a more intense UV field, but also some increase in the abundance of the
PAH and VSG grains. In the models, this would decrease the $k_{\rm B3V}$
values and probably improve the fits in the FIR regime where the observations 
appear to be more consistent with a lower large grain temperature.

In Sect. \ref{subsect:FIRopacity}, we derived 200\,$\mu$m optical depths of $1.4 \times 10^{-3}$
and $0.9 \times 10^{-3}$ for the ON1 and ON2 positions respectively. The
radiative transfer models including the B3V star radiation component
resulted in similar values, $\sim 1.5 \times 10^{-3}$ for ON1 and $\sim 0.8
\times 10^{-3}$ for ON2, and thus do not significantly alter the previous
$\sigma_{200}^{H}$ estimates that were $(1.9\pm0.6) \times
10^{-25}$\,cm$^{2}$ / H for ON1 and $(4.0\pm1.3) \times 10^{-25}$\,cm$^{2}$ / H
for ON2. The opacity is thus higher for the lower column density sightline
ON2. Both numbers are higher than the values $1.0 \times 10^{-25}
(\lambda / 250\,\mu{\rm m})$\,cm$^{2}$ / H derived for high latitudes
\citep{boulanger96b}. For the ON2 position, the value is actually similar
to what has been reported for apparently much denser clouds
(\citealt{hildebrand, beckwith1990, martin2012, lehtinen98}; see
Table 8 in Paper I).
It can be conjectured (for example) that photoevaporation caused by the
strong UV fields has decreased the grain sizes at the ON1 position. This
is not necessarily the case, however, because the dust opacities at both
positions are still higher than in diffuse medium and, because of the lower
opacity, ON2 should be even more susceptible to this process. It is more
likely that differences in volume density play a role. The ON2 position
lies at the centre of a filamentary structure and actually coincides with
a local column density peak that is visible in (for example) the $Herschel$
map (see Fig. 1). On the other hand, ON2 passes through a
envelope of LDN~1688 that is likely to be more extended along the
line of sight. In spite of the higher column density, the volume density
along the ON1 sightline could be lower than for ON2. In this case, the
difference in opacities could be explained by the growth of grain sizes in
a dense environment. Both sources have similarly high colour temperatures,
$T_{\rm d}\sim24$\,K, and the $A(J)$ values are too low to completely
shield the dust from the external UV field. Thus, these do not seem to be
suitable sites for the formation of either ice mantles or large dust
aggregates. In the case of ON2, however, the peak extinction could be
underestimated because $A(J)$ was derived from 2MASS stars at a
relatively low spatial resolution. In the radiative transfer models, the
dust temperatures were $\sim$20\,K or higher, but if the density
distribution is strongly peaked, the central extinction could be several
times higher. This would allow the existence of a significant amount of
dust below 15\,K without it having a large effect on the SEDs. 

\section{Conclusions}

Our ISOPHOT spectroscopic and spectrophotometric data for two
positions within LDN 1688 have probed the properties of the main dust
populations: PAH-like molecular species, transiently-heated VSGs and
large, classical grains.

The mid-infrared molecular emission band spectra are not markedly
different from those seen in other ISM environments, in spite of the
strong local radiation field. The 7.7\,$\mu$m / 11.3\,$\mu$m band
height ratio is $\sim$2.0 at both positions. This is lower than the
values observed toward G~300.2~-16.8.

As in Paper 1, a population of molecular species and/or VSGs is needed
to fit the 10\,$\mu$m continuum emission. The continuum is higher at
the position ON1 than at the position ON2. Taking into account the
high continuum levels in G~300.2~-16.8, this suggests that the
variation is due to dust properties, rather than a direct effect of a
high radiation field.

The strong local radiation field is reflected in the FIR colour
temperature that is $\sim$24\,K for both ON1 and ON2. However,
regarding the the MIR/FIR emission ratios, LDN 1688 is still similar
to the G~300.2~-16.8 ON3 position (Paper 1). The FIR opacity
($\tau_{200}$/$A(J)$ and $\sigma_{200}^{H}$) towards ON2 is
approximately twice as high as towards ON1.

With plausible assumptions of the relative positions of the stars and
the intervening extinction (see Appendix \ref{sect:ISRF_model2}), we
conclude that the stars Rho Oph AB and HD 147889 dominate the local
ISRF. The nearby Upper Sco OB association has a smaller contribution,
mainly at the ON2 position. The total radiation field intensity is probably
at least one order of magnitude above the value in the solar neighbourhood.

In radiative transfer models, the spectra at ON1 and ON2 can be fitted
by increasing the local ISRF by a factor of $\sim$20 -- 30 and the
abundance of PAHs by a factor of a few.  However, the modelling also
reveals a strong degeneracy between the assumptions of the radiation
field and the dust properties. The best fits are obtained by taking
into account the harder radiation originating in the B-type stars,
almost without any modification of the PAH abundance. However, all
models underestimate the 10-$\mu$m continuum.

The small differences between the sightlines can be understood by the
higher optical depth of the sightline ON1 counter-balancing the higher
intensity of the radiation field. The higher dust opacity towards ON2,
in spite of the lower column density, could be related to higher
volume density along that line-of-sight.

\section*{Acknowledgments}
The authors wish to warmly thank the anonymous referee for the very thorough and 
constructive criticism which led to substantial improvements to this paper.
The authors would like to thank Fran\c{c}ois Boulanger for providing us with the
correctly-calibrated ISOCAM-CVF datacube and Peter \'Abrah\'am for use
of his ZL template spectra. MGR gratefully acknowledges support from the Finnish
Academy (grant No. 174854), the Magnus Ehrnrooth Foundation, the
Joint Astronomy Centre, Hawaii (UKIRT), and the ALMA project. MJ acknowledges
support by Academy of Finland grants 127015 and 250741. ISOPHOT and the Data
Centre at MPIA, Heidelberg, are funded by the Deutsches Zentrum f\"ur Luft- und
Raumfahrt DLR and the Max-Planck-Gesellschaft. KL acknowledges support by
Academy of Finland grant 132291. DL is indebted to DLR, Bonn, the
Max-Planck-Society and ESA for supporting the {\it ISO} Active Archive Phase.
This research has made use of NASA's Astrophysics Data System. This
research has made use of the SIMBAD database, operated at CDS, Strasbourg,
France. This publication makes use of data products from the Wide-field Infrared Survey Explorer,
which is a joint project of the University of California, Los Angeles, and the Jet Propulsion
Laboratory/California Institute of Technology, funded by the National Aeronautics and
Space Administration.

\appendix

\section{The Interstellar Radiation Field in LDN~1688}
\label{sect:ISRF_model2}

The local ISRF in LDN~1688 is dominated by two B-type double stars. 
Their distances, spectral types, luminosities, and extinctions are:\\
\\
 $\rho$ Oph AB: $d=111^{+12}_{-10}$ pc, B2V+B2V, 5800 L$_{\sun}$, $A(J) = 0.60$ mag,       \\
 HD 147889 AB: $d=118^{+13}_{-11}$ pc, B2IV+B3IV, 7800 L$_{\sun}$, $A(J) = 1.45$ mag.       \\    
 
The distances are from the revised HIPPARCOS catalogue of \cite{vanleeuwen2007}, spectral types from SIMBAD and \citet{casassus},
corresponding luminosities from \citet{dejager}, and extinctions based on \citet{carrasco,whittet1974}.

Using VLBA parallax measurements of two embedded stars, DoAr21 and  S1, 
located close to HD 147889, \citet{loinard2008} derived the distance  $d = 120.0^{+4.5}_{-4.2}$ pc, 
which we adopt here as the distance to LDN~1688. 

The extinctions through the cloud at the sightlines of the stars (cf. Sect. 2.4),  
$A(J) = 0.88 \pm 0.05$ for $\rho$ Oph AB and $A(J)=2.88\pm 0.07$ for HD~147889, 
are clearly larger than the extinctions of the stars. 
$\rho$ Oph AB and HD147889 are surrounded by bright and extended blue reflection nebulae,
vdB 105 and vdB 106 (cf. Fig. 1 of \citealt{wilking2008}). We conclude that both  $\rho$ Oph AB and HD147889 
must be embedded in the cloud. Their distances are compatible with that of LDN~1688,
i.e. 120 pc. Assuming that the stars and the dust in ON1 and ON2 are at the same distance of 120 pc
from the Earth (i.e. all are in the plane of sky) we obtain the following values for their separations:

$\rho$ Oph AB to ON1 1.44 pc, to ON2 0.97 pc;

HD147889 to ON1 0.75 pc, to ON2 3.05 pc.\\
The presence and extension of the blue reflection nebulae also demonstrate that ON1 and ON2 are exposed 
to substantial optical illumination from $\rho$ Oph AB. ON1 is exposed to illumination mainly from HD~147889.

In order to estimate more quantitatively the contribution to the ISRF by the stars, we make use of
results by \citet{habing} for the ISRF at  $\lambda = 1000-2200$ \AA.   For different spectral types,
Table 3 of \citet{habing} lists the distance, $r_{100}$, from a star at which the stellar contribution to the
radiation density equals the average Solar neighbourhood ISRF value
between  $\lambda = 1000-2200$ \AA\, (often denoted by $G_0 = 1$).
For $\rho$ Oph AB and HD~147889, $r_{100} = 8.5$ pc and 9.6 pc, 
respectively.  We thus find the following ISRF contributions in
Habing units for the case of no intervening extinction:

$\rho$ Oph AB at ON1 $G_0 = 35$, at ON2 $G_0 = 77$;

HD147889 at ON1  $G_0 = 164$, at ON2 $G_0 = 10$.

For an estimate of the {\it intervening} extinction at 2000 \AA\,  influencing the light passing from each of
the stars to position ON1 or ON2, we adopt, somewhat arbitrarily, half of the line-of-sight extinction
through the cloud: for $\rho$ Oph AB  $A(J) = 0.44$ both for ON1 and ON2, and 
for HD~147889 $A(J) = 1.44$ for ON2 but, because of the shorter traversed distance,
$A(J) = 0.48$ (i.e. $\sim$ 1/3 of the full-distance value) for ON1. 
These values correspond to a model in which the cloud's optical depth is transversely as large as 
along the line of sight. 
For the transformation from $A(J)$ to $A$(2000 \AA\,) we use 
the extinction parameter value $R_V = A(V) / E(B-V)= 4.2$, valid for the $\rho$ Oph cloud \citep{whittet}.
Part of the stellar light lost by extinction is returned to the forward direction by scattering.
For complete forward scattering with albedo $a$, the effective extinction is $A_{eff} = A \times (1-a)$. With $a\approx$ 0.4 and fairly strong forward scattering with $g \approx 0.6$ 
at 2000 \AA\, \citep{li01b}, a substantial part of the scattering is not directed forwards. We nevertheless use
the estimate $A_{eff} = A \times (1-a) = 0.6 \times A$ which is motivated by the phenomenon that the 
penetration of scattered light is substantially favoured by a clumpy medium.  
The effective extinctions $A$(2000 \AA\,) are thus:

$\rho$ Oph AB to ON1 and ON2 1.7 mag;

HD147889 to ON1 1.8 mag, to ON2 5.8 mag.

With these extinction corrections the the ISRF contributions by the two stars obtain the following estimates:

$\rho$ Oph AB at ON1  $G_0 = 7$ , at ON2  $G_0 = 16$;
HD147889 at ON1  $G_0 = 28$, at ON2 $G_0 = 0.05$.

These values refer to the wavelength region 1000 - 2200 \AA\, \citep{habing}. 
The corresponding $k_{B3V}$ values referring to the total UV-to-near-IR ISRF
as defined in Sect. 4, and added up for contributions from both stars, are the following:

 $k_{B3V}$(ON1) = 6.2 and  $k_{B3V}$(ON2) = 2.7.

The uncertainty is at least factor of 2
and is caused mainly by the uncertainty in the intervening extinction estimate. With no
intervening extinction the values would be:

$k_{B3V}$(ON1) = 26 and  $k_{B3V}$(ON2) = 11.

The radiation field estimates obtained in Sect. \ref{sect:RTmodels} (Figs. ~\ref{fig:ON1_LD} and
~\ref{fig:ON2_LD}, lower panels) are larger by a factor of 3 -- 4 when extinction is included. For
the (unrealistic) case with no extinction, the agreement would be within $\sim$ 30 per cent. 

Besides $\rho$ Oph AB and  HD~147889, other members of the Upper Scorpius OB association
also contribute to the radiation field in LDN~1688. With a distance of 144 pc to the centroid 
($l \sim 350$ deg, $b \sim 20$ deg) of the early type stars in the association \citep{dezeeuw}
and using the radiation 
parameter $r_{100} = 67$ pc as given in \citet{habing} we find a contribution of $G_0 \sim 7$
for no intervening extinction. Because the treatment 
with a point source at its centroid is not a good approximation for the widely distributed association
of stars, we have also calculated separately the contributions of the following eight dominating
early B type stars: $\zeta$ Oph, $\chi$ Oph, $\nu$ Sco A, $\omega$ Sco, $\delta$ Sco, $\pi$ Sco,
$\tau$ Sco, $\lambda$ Lib \citep{sujatha}. The result is $G_0 = 1.9$ for ON1 and $G_0 = 2.2$ for
ON2, the largest contribution  of $G_0 \sim 0.7$ coming from the O9~V type star $\zeta$ Oph. 
Thus, the contribution by  the Upper Scorpius OB association falls clearly below
that of $\rho$ Oph AB and HD~147889 but may be relevant for ON2 which is close to
the northern surface of LDN~1688 at which most of the radiation from the association is impinging.

We note that \citet{laureijs1995} have estimated the contribution of Upper Sco OB
to the radiation field impinging the L134/183 dark cloud complex, located on the northern side 
(at $l \sim4$ deg, $b \sim36$ deg, $d = 110$ pc) of the association but further away from it 
($\sim60$ pc) than LDN 1688. Using Habing's (1968) results they found a contribution 
at 1000 -- 2200 \AA\, corresponding to  $G_0 \sim 2$. 

The cluster of newly-born stars \citep{wilking2008} embedded in the dense molecular core $\sim10$' West
of HD~147889 is also a potential source of additional ISRF, at least for position ON1. Its near- and mid-IR
radiation could be sufficient to contribute to the heating of the big grains.
Using $COBE$ / DIRBE maps we determined the mean surface brightness of the cluster 
at 2 -- 240 $\mu$ m within a circle of $0.7^{\circ}$ radius.
An observer located at position ON1 would see this same surface brightness over about
half the sky (2$\pi$ steradians). When multiplied with the absorption coefficient of the \citet{li01b} dust
model, we find that the total absorbed energy from this radiation source corresponds to $\sim$ half of
the energy absorbed from an isotropic \citet{mathis83} ISRF (i.e. $k_{ISRF}=1.0$).
We therefore conclude that this radiation field component is not important for the dust heating in LDN~1688.

\label{lastpage}


\begin{thebibliography}{99}

\bibitem[\protect\citeauthoryear{Bakes et al.}{2001}]{bakes2001}Bakes E. L. O., Tielens A. G. G. M., Bauschlicher C. W., Jr., Hudgins, D. M., Allamandola, L. J., 2001, ApJ, 560, 261
\bibitem[\protect\citeauthoryear{Bernard, Boulanger \& Puget}{1993}]{bernard93}Bernard J. -P., Boulanger F., Puget J. L., 1993, A\&A, 277, 609
\bibitem[\protect\citeauthoryear{Bontemps et al.}{1996}]{bontemps1996}Bontemps S., Andr{\'e} P., Tereby S., Cabrit S., 1996, A\&A, 311, 858 
\bibitem[\protect\citeauthoryear{Beckwith et al.}{1990}]{beckwith1990}Beckwith, S. V. W., Sargent, A. I., Chini, R. S., Guesten, R., 1990, AJ, 99, 924
\bibitem[\protect\citeauthoryear{Boulanger et al.}{1996a}]{boulanger96a}Boulanger F., et al., 1996, A\&A, 315, L325
\bibitem[\protect\citeauthoryear{Boulanger et al.}{1996b}]{boulanger96b}Boulanger F., et al., 1996, A\&A, 312, 256
\bibitem[\protect\citeauthoryear{Boulanger et al.}{2005}]{boulanger05}Boulanger F. et al., 2005, A\&A, 436, 1151
\bibitem[\protect\citeauthoryear{Bruzual \& Charlot}{2003}]{bruzual2003}Bruzual G., Charlot, S., 2003, MNRAS, 344, 1000
\bibitem[\protect\citeauthoryear{Carrasco, Strom \& Strom}{1973}]{carrasco} Carrasco L., Strom S.~E., Strom K.~M., 1973, ApJ, 182, 95 
\bibitem[\protect\citeauthoryear{Casassus et al.}{2008}]{casassus} Casassus S., et al., 2008, MNRAS, 391, 1075 
\bibitem[\protect\citeauthoryear{Chan et al.}{2001}]{chan2001}Chan, K.-W., Roellig, T. L., Onaka, T., Mizutani, M., Okumura, K., Yamamura, I., Tanab{\'e}, T., Shibai, H., Nakagawa, T., Okuda, H., 2001, ApJ, 546, 273
\bibitem[\protect\citeauthoryear{del Burgo et al.}{2003}]{delBurgo}del Burgo C., Lauerijs R. J., \'Abrah\'am P., Kiss, Cs., 2003, MNRAS, 346, 403
\bibitem[\protect\citeauthoryear{de Jager \& Nieuwenhuijzen}{1987}]{dejager} de Jager C., Nieuwenhuijzen H., 1987, A\&A, 177, 217 
\bibitem[\protect\citeauthoryear{de Zeeuw et al.}{1999}]{dezeeuw} de Zeeuw P.~T., Hoogerwerf R., de Bruijne J.~H.~J., Brown A.~G.~A., Blaauw A., 1999, AJ, 117, 354
\bibitem[\protect\citeauthoryear{Gabriel}{2000}]{gabriel}Gabriel C., 2000, PHT Interactive Analysis User Manual (V9.0) \\
www.iso.vilspa.esa.es/manuals/PHT/pia/um/pia\_um.html
\bibitem[\protect\citeauthoryear{Galliano et al.}{2008}]{galliano2008}Galliano F., Madden S. C., Tielens A. G. G. M., Peeters E., Jones A. P., 2008, ApJ, 679, 310
\bibitem[\protect\citeauthoryear{Habing}{1968}]{habing} Habing H.~J., 1968, BAN, 19, 421
\bibitem[\protect\citeauthoryear{Hildebrand}{1983}]{hildebrand}Hildebrand R. H., 1983, QJRAS, 24, 267
\bibitem[\protect\citeauthoryear{Kahanp\"a\"a et al.}{2003}]{kahanpaa2003}Kahanp\"a\"a J., Mattila K., Lehtinen K., Leinert C., Lemke D., 2003, A\&A, 405, 999
\bibitem[\protect\citeauthoryear{Kessler et al.}{1996}]{kessler}Kessler M. F. et al., 1996, A\&A, 315, L27
\bibitem[\protect\citeauthoryear{Klaas et al.}{1994}]{klaas94}Klaas U., Kr\"uger H., Heinrichsen I., Heske A., Laureijs, R., eds, 1994, ISOPHOT Observers Manual, version 3.1.1
\bibitem[\protect\citeauthoryear{Klaas et al.}{2002}]{klaas02}Klaas U. et al., 2002, ISOPHOT Calibration Accuracies, Version 5.0, SAI/1998-092/Dc
\bibitem[\protect\citeauthoryear{Laureijs et al.}{1995}]{laureijs1995} Laureijs R.~J., Fukui Y., Helou G., Mizuno A., Imaoka K., Clark F.~O., 1995, ApJS, 101, 87 
\bibitem[\protect\citeauthoryear{Lehtinen et al.}{1998}]{lehtinen98}Lehtinen K., Lemke D., Mattila K., Haikala L.K., 1998, A\&A, 333, 702
\bibitem[\protect\citeauthoryear{Leinert et al.}{2002}]{leinert}Leinert C., \'Abrah\'am P., Acosta-Pulido J., Lemke D., Siebenmorgen R., 2002, A\&A, 393, 1073
\bibitem[\protect\citeauthoryear{Lemke et al.}{1996}]{lemke96}Lemke D. et al., 1996, A\&A 315, L64
\bibitem[\protect\citeauthoryear{Li \& Draine}{2001}]{li01b}Li A., Draine, B. T., 2001, ApJ, 554, 778
\bibitem[\protect\citeauthoryear{Li \& Greenberg}{1997}]{li97}Li A., Greenberg J. M., 1997, A\&A, 323, 566
\bibitem[\protect\citeauthoryear{Loinard et al.}{2008}]{loinard2008}Loinard, L., Torres, R. M., Mioduszewski, A. J., Rodr'guez, L. F., 2008, ApJ, 675, L29
\bibitem[\protect\citeauthoryear{Lombardi, Lada \& Alves}{2008}]{lombardi2008}Lombardi M., Lada C. J., Alves J., 2008, A\&A, 489, 143
\bibitem[\protect\citeauthoryear{Lombardi \& Alves}{2001}]{lombardi}Lombardi M., Alves J., 2001, A\&A, 377, 1023
\bibitem[\protect\citeauthoryear{Martin et al.}{2012}]{martin2012}Martin, P. G. et al., 2012, ApJ, 751, 28 
\bibitem[\protect\citeauthoryear{Mathis}{1990}]{mathis90} Mathis J.~S., 1990, ARA\&A, 28, 37
\bibitem[\protect\citeauthoryear{Mathis}{1996}]{mathis1996}Mathis J. S., 1996, ApJ, 472, 643
\bibitem[\protect\citeauthoryear{Mathis, Mezger \& Panagia}{1983}]{mathis83}Mathis J. S., Mezger P. G., Panagia N., 1983, A\&A, 128, 212
\bibitem[\protect\citeauthoryear{Padgett et al.}{2008}]{padgett2008}Padgett D. L. et al., 2008, ApJ 672, 1013
\bibitem[\protect\citeauthoryear{Pilbratt et al.}{2010}]{pilbratt2010}Pilbratt, G. L., Riedinger, J. R., Passvogel, T., Crone, G., Doyle, D., Gageur, U., Heras, A. M., Jewell, C., Metcalfe, L., Ott, S., Schmidt, M., 2010, A\&A, 518, L1
\bibitem[\protect\citeauthoryear{Poglitsch et al.}{2010}]{poglitsch2010}Poglitsch, A. et al., 2010, A\&A, 518, L2
\bibitem[\protect\citeauthoryear{Puget \& L\'{e}ger}{1989}]{puget}Puget J. L., L\'{e}ger A., 1989, ARA\&A, 27, 161
\bibitem[\protect\citeauthoryear{Rawlings et al.}{2005}]{rawlings2005}Rawlings, M. G., Juvela M., Mattila K., Lehtinen K., Lemke D., 2005, MNRAS 356, 810, Paper 1
\bibitem[\protect\citeauthoryear{Sellgren}{1984}]{sellgren}Sellgren K., 1984, ApJ, 277, 623
\bibitem[\protect\citeauthoryear{Sujatha et al.}{2005}]{sujatha} Sujatha N.~V., Shalima P., Murthy J., Henry R.~C., 2005, ApJ, 633, 257 
\bibitem[\protect\citeauthoryear{Tielens}{2008}]{tielens2008}Tielens, A. G. G. M., 2008, ARA\&A, 46, 289
\bibitem[\protect\citeauthoryear{van Leeuwen}{2007}]{vanleeuwen2007} van Leeuwen F., 2vanleeuwen007, A\&A, 474, 653
\bibitem[\protect\citeauthoryear{Vuong et al.}{2003}]{vuong2003}Vuong M. H., Montmerle T., Grosso N., Feigelson E. D., Verstraete L., Ozawa H., 2003, A\&A 408, 581
\bibitem[\protect\citeauthoryear{Werner et al.}{2004}]{werner2004}Werner, M. et al., 2004, ApJS, 154, 1
\bibitem[\protect\citeauthoryear{Whittet}{1974}]{whittet1974} Whittet D.~C.~B., 1974, MNRAS, 168, 371 
\bibitem[\protect\citeauthoryear{Whittet et al.}{2001}]{whittet}Whittet, D. C. B., Gerakines, P., Hough, J. H., Shenoy, S. S., 2001, ApJ, 547, 872
\bibitem[\protect\citeauthoryear{Wilking, Gagn\'{e} \& Allen}{2008}]{wilking2008}Wilking, B. A., Gagn\'e, M., Allen, L. E., ed., Handbook of Star Forming Regions, Volume II: The Southern Sky. Astron. Soc. Pac., San Francisco, p. 351
\bibitem[\protect\citeauthoryear{Wright, E. L. et al.}{2010}]{wright2010}Wright E. L. et al., 2010, AJ, 140, 1868

\end{thebibliography}
\end{document}